# Elastic, thermodynamic, electronic and optical properties of recently discovered transition metal boride NbRuB superconductors: an *ab-initio* investigation

F. Parvin and S. H. Naqib[*]

Department of Physics, University of Rajshahi, Rajshahi 6205, Bangladesh

**Abstract**

The elastic, thermodynamic, electronic, and optical properties of recently discovered and potentially technologically important transition metal boride NbRuB, have been investigated using the density functional formalism. Both generalized gradient approximation (GGA) and local density approximation (LDA) were used for geometrical optimization and for estimation of various elastic moduli and constants. The optical properties of NbRuB have been studied for the first time with different photon polarizations. The frequency (energy) dependences of various optical constants compliment quite well to the essential features of the electronic band structure calculations. Debye temperature of NbRuB has been estimated form the thermodynamical study. All these theoretical estimates have been compared with published results, where available, and discussed in detail. Both electronic band structure and optical conductivity reveal robust metallic characteristics. NbRuB possesses significant elastic and electronic anisotropy. The effects of electronic band structure and Debye temperature on the emergence of superconductivity have also been analyzed.

## 1. Introduction

Ternary and binary borides (e.g., diborides) belong to a remarkable class of materials mainly because of their variety of attractive physical properties suitable for diverse fields of industrial applications. One of the common features of these transition metal borides is their high value of hardness [1].

[*] Corresponding author. Email: salehnaqib@yahoo.com

Besides, transition metal borides show variety of interesting electronic ground states, including superconductivity and magnetic order [2]. One of the most prominent examples is $Nd_2Fe_{14}B$ – the hardest among all the ferromagnetic compounds [3]. Almost all of the metallic borides show high level of hardness and particularly the binary diborides like $RuB_2$ and $OsB_2$ are considered to be superhard and ultraincompressible [1]. $ReB_2$ is expected to be able to scratch even diamond [4]. Transition metal borides often show refractory behavior and chemical inertness suitable for heavy duty coarse condition applications. Attractive electronic band structure related features in some of the metallic borides (e.g., significant optical conductivity, reasonable electronic density of states at the Fermi level, non-selective and high value of reflectivity etc.) [5] make them suitable for variety of potential electronic and optoelectronic applications. Materials with high hardness are used under extreme pressure and temperature conditions. Their applications as cutting tools and in hard coating are well documented [6]. Therefore, study and understanding of the thermo-mechanical properties of these transition metal borides are of substantial importance. In addition, various transition metal ternary borides (e.g., Mg (or V, Cr, Mn, Co) borides) have significant prospect for being used as hydrogen storage material [7]. Electrically conducting (metallic) boride compounds not only gives rise to the formation of one-, two-, or three-dimensional arrangements of covalently bonded boron atoms, but also, due to intricate structural features, gives rise to a multitude of electronic interactions that can lead to superconductivity or various types of magnetic orders [8 – 23]. In recent years Nb-Ru-B systems have attracted considerable interest of the materials science community [24].

Till now, only four types of Nb-Ru-B ternary borides had been reported and investigated, namely, $Nb_7Ru_6B_8$, $Nb_3Ru_5B_2$, $Nb_2RuB_2$ and NbRuB [2, 24 – 27]. The arrangement of the boron atoms in these compounds depends strongly on the metal element-boron (M/B) ratio. A number of different structural schemes are found with boron substructures ranging from isolated boron atoms to boron fragments or chains [28]. A special type of these phases has been the subject of several research groups – borides containing a sheet arrangement of atoms that leads to the separation of isolated boron atoms in one layer and boron fragments in the other, with metal atoms present in each layer. Recently, Mbarki et al. [24] succeeded in synthesizing NbRuB transition metal ternary boride by arc-melting of the elements in water-cooled crucibles kept in argon atmosphere. This compound shows a layered structure with an orthorhombic space group *Pmma*. NbRuB consists of two layers, one containing the Nb and an isolated B atoms and the other with Ru atoms and boron fragments (dumbbell shaped $B_2$ cluster).

A number of theoretical studies have been carried out Nb-Ru-B systems [2, 24, 27, 29 – 31]. Mbarki et al. [24] studied the electronic structure and the bonding properties of NbRuB using the Generalized Gradient Approximation/Linear Muffin Tin-Orbital (GGA-LMTO) formalism. This study predicted metallic behavior for NbRuB with strongest bonding between the B atoms in the dumbbell situated in the same layer. *Ab-initio* investigation of the structural and elastic properties of predicted $Nb_2MB_2$ (M = Tc, Ru, Os) phases under hydrostatic pressure were done by Li et al. [31]. Similar ambient pressure properties were studied by the same group [31]. Fokwa et al. [29] theoretically examined the influence of bonding properties and electronic structure on the bulk and shear modulus of $A_2MB_2$ series (A = Nb, Ta and M = Fe, Ru, Os).

Very few theoretical studies exist for the recently synthesized NbRuB compound. Beside the study by Mbarki et al. [24], Tian and Chen [30] focused mainly on thermodynamic and superconducting state properties of this ternary transition metal boride system. To the best of our knowledge, no study of optical properties exists in the literature. A comprehensive study of thermodynamical properties is also lacking. In this study we wish to fill these voids. Investigation of optical properties compliments the electronic band structure calculations. Photon polarization dependent studies can reveal valuable information regarding any underlying electronic anisotropy. Moreover, these hard transition metal ternary borides may have potential optoelectronic utilities. To explore this possibility optical study is essential. In this paper we have we have studied the structural, elastic, thermal, electronic and optical properties of NbRuB using the Density Functional Theory (DFT). Pressure and temperature dependences of the bulk modulus and the volume thermal expansion coefficient have also been investigated. We have also used the quasi-harmonic Debye approximation to explore the behavior of Debye temperature and specific heat of NbRuB at various pressures and temperatures.

The paper is organized as follows. In Section 2 the computational methodology is described in brief. Various theoretical results are presented and analyzed in Section 3. Finally, the results are discussed and important conclusions are drawn in Section 4.

## 2. Computational Methodology

The *ab-initio* calculations presented in this paper were carried out using the Cambridge Serial Total Energy Package (CASTEP) code [32]. CASTEP employs the plane wave pseudopotential approach based on the density functional theory [33, 34]. During the first-principles calculations the choice of exchange-correlation potential is important. In this study we have used both LDA (with Ceperley-Alder (CA) and Perdew-Zunger (PZ) functional) and GGA exchange correlation functional. LDA is good enough for structural, elastic and vibrational properties in many cases. LDA, though has a tendency of underestimating the lattice constants and overestimating various elastic constants and moduli in certain compounds. GGA is more reliable and realistic since it permits a variation in the electron density. Here we have applied GGA functional as parameterized by the Perdew-Burke-Ernzerhof (PBE) scheme [35]. GGA relaxes the lattice constants due to the repulsive core-valence electron exchange correlation. Vanderbilt-type ultrasoft pseudopotentials were used to model the electron-ion interactions. This relaxes the norm-conserving criteria but produces a smooth and computation friendly pseudopotential. A $4 \times 13 \times 7$ $k$-point mesh of Monkhorst-Pack scheme [36] was employed to sample the first Brillouin zone. During the calculations, a plane-wave cutoff kinetic energy of 500 eV was used to limit the number of plane waves within the expansion. For geometrical optimization, the crystal structures were fully relaxed by the widely employed Broyden-Fletcher-Goldfrab-Shanno (BFGS) scheme [37]. The convergence tolerances used for various parameters are: energy = $1.0 \times 10^{-5}$ eV/atom; maximum force = 0.03 eV/Å; maximum stress = 0.05 GPa; maximum displacement = 0.001 Å; self consistent field = $1.0 \times 10^{-6}$ eV/atom.

The elastic constants and moduli have been obtained by applying a set of finite homogeneous crystal deformations and calculating the resulting stresses, as implemented within the CASTEP code.

The frequency dependent optical properties of a compound are extracted from the estimated dielectric function, $\varepsilon(\omega) = \varepsilon_1(\omega) + i\varepsilon_2(\omega)$, which describes the interactions of photons with electrons. The imaginary part, $\varepsilon_2(\omega)$, of the dielectric function is calculated in the momentum representation of matrix elements between occupied and unoccupied electronic quantum states by employing the CASTEP supported formula given by

$$\varepsilon_2(\omega) = \frac{2e^2 \pi}{\Omega \varepsilon_0} \sum_{k,v,c} \left| \langle \psi_k^c | \hat{u} \cdot \vec{r} | \psi_k^v \rangle \right|^2 \delta(E_k^c - E_k^v - E) \quad (1)$$

Here, $\Omega$ is the unit cell volume, $\omega$ frequency of the incident photon, $e$ is the charge of an electron, $\hat{u}$ is the unit vector defining the incident electric field polarization, and $\psi_k^c$ and $\psi_k^v$ are the conduction and valence band wave functions at a given wave-vector $k$, respectively. The Kramers-Kronig transformations yield the real part $\varepsilon_1(\omega)$ of the dielectric function from the corresponding imaginary part $\varepsilon_2(\omega)$. The refractive index, the absorption spectrum, the energy loss-function, the reflectivity, and the optical conductivity (the real part) can be derived once $\varepsilon_1(\omega)$ and $\varepsilon_2(\omega)$ are known [38]. The intraband electronic contribution to the optical properties of metallic systems affects mainly the low-energy part of the spectra. This can be corrected for the dielectric function by including an empirical Drude term with unscreened plasma frequency of 6 eV and a damping term of 0.05 eV.

It should be mentioned that, for investigation of thermodynamic properties, we have used the energy-volume data calculated from the third-order Birch-Murnahgan equation of state [39] using the zero temperature and zero pressure equilibrium values of energy, volume, and bulk modulus obtained through DFT calculations.

## 3. Results and Analysis

### 3.1 Structure of NbRuB

The optimized crystal structure (space group *Pmma*) of NbRuB is shown in Fig. 1. The structural parameters are listed and compared with available experimental and theoretical results in Table 1.

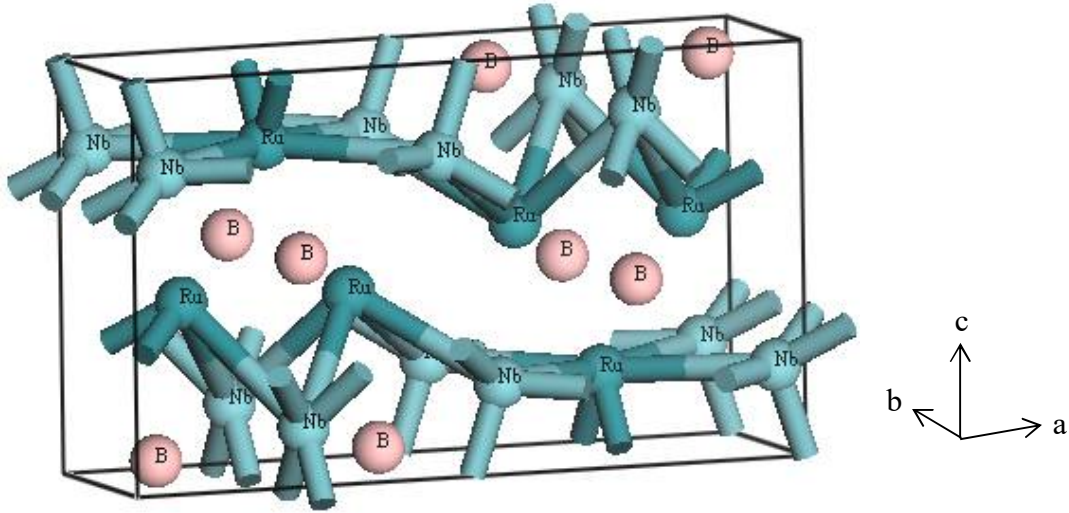

Figure 1: Crystal structure of NbRuB. The layers with isolated boron atoms and boron dumbbells are shown.

Table 1: Optimized lattice parameters ($a$, $b$, $c$) and unit cell volume ($V$) of NbRuB.

| $a$(Å) | $b$(Å) | $c$(Å) | $V$(Å$^3$) | Reference |
|---|---|---|---|---|
| 10.780 | 3.115 | 6.298 | 211.51 | LDA[this study] |
| 10.838 | 3.137 | 6.331 | 215.23 | GGA[this study] |
| 10.833 | 3.141 | 6.324 | 215.14 | [30] GGA[theoretical] |
| 10.867 | 3.156 | 6.350 | 217.67 | [27][experimental] |
| 10.869 | 3.165 | 6.350 | 218.47 | [24][experimental] |

From Table 1 it is seen that the equilibrium lattice parameters obtained with GGA agree quite well with the experimental values. GGA underestimates the lattice parameters slightly. But we should keep in mind that the experimental lattice constants were extracted at room temperature while the *ab-initio* estimates were done at absolute zero. On the other hand LDA grossly underestimates the lattice parameters. This is somewhat expected due to the localized nature of the LDA functional.

The structure of NbRuB consists of two different layers stacked alternately along the crystallographic *c*-direction. The first layer (the bottom one in Fig. 1) has Nb and B atoms, while the second one is filled with Ru and $B_2$ dumbbells in addition to Nb. In NbRuB we can identify two different types of B atom, one is situated at the center of a triangular prism with six Ru at vertices and the other as $B_2$ dimer inside a double-triangular prism constructed with Nb atoms.

*3.2 Elastic properties*

Table 2 exhibits the various elastic constants of NbRuB under ambient condition.

Table 2. Independent single crystal elastic constants, $C_{ij}$, (all expressed in GPa) of NbRuB at $P = 0$.

| Compound | $C_{11}$ | $C_{22}$ | $C_{33}$ | $C_{44}$ | $C_{55}$ | $C_{66}$ | $C_{12}$ | $C_{13}$ | $C_{23}$ | Reference |
|---|---|---|---|---|---|---|---|---|---|---|
| NbRuB | 521.1 | 410.7 | 525.6 | 215.1 | 180.6 | 198.9 | 241.5 | 165.8 | 257.2 | LDA[this study] |
| | 507.7 | 377.1 | 512.6 | 206.3 | 178.1 | 192.6 | 233.2 | 165.8 | 248.6 | GGA[this study] |
| | 523.2 | 361.9 | 504.8 | 212.6 | 181.8 | 196.2 | 217.2 | 154.9 | 259.3 | [30] GGA |

Table 3 shows the estimated elastic moduli of NbRuB.

Table 3. Bulk modulus $B$, Shear modulus $G$, Young's modulus $Y$, $B/G$ ratio, Poisson's ratio $v$, elastic anisotropy factor $A$, and the pressure derivative of bulk modulus $B'$ of NbRuB.

| $B$ (GPa) | $G$ (GPa) | $Y$ (GPa) | $B/G$ | $v$ | $A$ | $B'$ | Reference |
|---|---|---|---|---|---|---|---|
| 309.4 | 162.7 | 415.3 | 1.90 | 0.276 | 1.54 | 4.0 | LDA[this study] |
| 298.3 | 155.4 | 397.2 | 1.92 | 0.278 | 1.50 | 3.3 | GGA[this study] |
| 293.4 | 155.8 | 397.2 | 1.88 | 0.270 | - | - | [30] GGA |

The single crystal elastic constants and the elastic moduli shown in Tables 2 and 3 completely determine the mechanical response of NbRuB under applied external stress. The anisotropy factor $A$ was obtained from the relation $A = 2C_{44}/(C_{11} - C_{12})$. It is seen that NbRuB possesses significant elastic anisotropy. For an isotropic compound $A = 0$. There is another measure of elastic anisotropy, $k_c/k_a = (C_{11} + C_{12} - 2C_{13})/(C_{33} - C_{13})$. $k_c/k_a$ for NbRuB is 1.20. $k_c$ and $k_a$ denote the compressibility along $c$ and $a$ directions, respectively. This implies that the compressibility of NbRuB along $c$-direction is larger than that within the *ab*-plane. LDA estimates of elastic moduli are somewhat larger than those from GGA, which is expected.

Calculated elastic constants, $C_{ij}$, fulfill the various inequality criteria for mechanical stability [40]. The bulk and shear moduli of NbRuB is greater than that for Nb$_2$RuB$_2$ (B = 272 GPa and G = 146 GPa) [25]. The difference between single crystalline elastic constants, ($C_{12} - C_{44}$) is well known as the Cauchy pressure [41]. A positive value of ($C_{12} - C_{44}$) indicates the metallic bonding, whereas a negative value signifies covalent bonding. Based on this scheme, we find that NbRuB has predominantly metallic bonding in its structure. A positive Cauchy pressure always indicates ductile nature of a material, while a negative value corresponds to brittleness. Hence the ternary boride NbRuB should behaves in a ductile manner. Bulk modulus of NbRuB is also larger than that of Nb$_2$TcB$_2$ (284 GPa). Nb$_2$OsB$_2$, on the other hand, has a larger bulk modulus (estimated to be 354 Gpa) [31] compared to NbRuB. Both bulk modulus and the shear modulus give rough estimate of the hardness of a compound. Tian and Chen have estimated the hardness of NbRuB [30] about 15 GPa. This corresponds quite well with the results obtained in this work in the light of the findings reported in Ref. [6]. The parameter, $B/G$, is known as the Pugh's ratio [42] which distinguishes between ductile and brittle response of a compound under applied stress. The critical value of Pugh's ratio is estimated to be 1.75. A larger value indicates ductile behavior. Another criterion often used to differentiate between ductile and brittle behavior is the Poisson's ratio, $v$. A Poisson's ratio greater than 0.26 implies ductile behavior. Therefore, both Pugh's ratio and Poisson's ratio predict moderately ductile response to stress for NbRuB ternary boride. All these support the predictions from Cauchy pressure analysis.

*3.3 Thermal and pressure dependent properties*

Hard and superhard transition metal borides are often used under extreme conditions of pressure and temperature. Therefore, information regarding the response of the bulk modulus, coefficient of thermal expansion, etc. at various elevated pressures and temperatures is important. Fig. 2a shows the variation of the bulk modulus with pressure. The bulk modulus of NbRuB increases quite sharply with increasing hydrostatic pressure, consistent with the high calculated value of $B'$ (Table 3). The temperature dependent bulk modulus is shown in Fig. 2b. It is seen that at low temperature, the bulk modulus remains almost constant, a standard behavior related to the third law of thermodynamics [43]. A quasi-linear decrease of the bulk modulus with at higher temperature is observed.

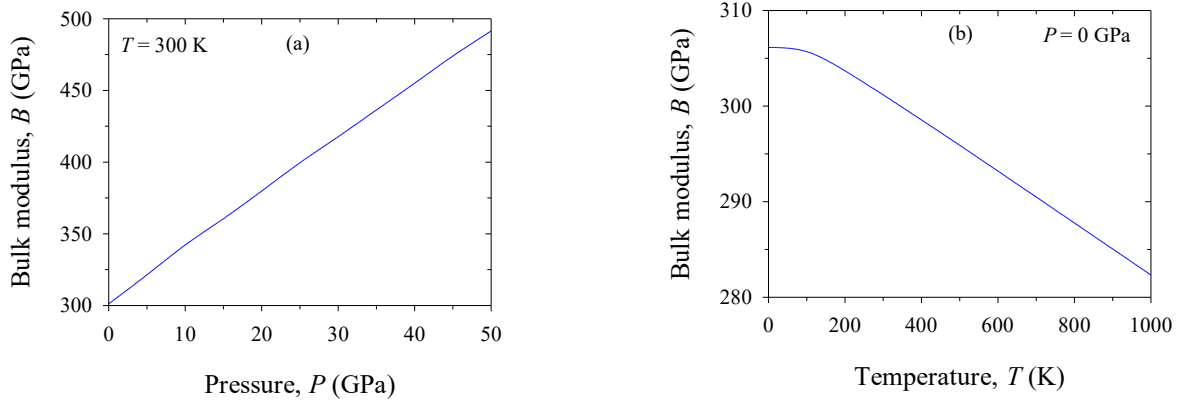

Figure 2: (a) Pressure and (b) Temperature dependences of the bulk modulus of NbRuB.

The temperature and pressure dependent volume thermal expansion coefficient, $\alpha_V$, are shown in Figs. 3a and 3b, respectively. Like all the elastic constants, volume thermal expansion coefficient is also directly related to the bonding properties of a solid. The melting point of a solid often shows an inverse relationship with $\alpha_V$. $\alpha_V$ of NbRub increases sharply at low temperatures but flattens above 400 K. $\alpha_V$ decreases with applied pressure. The rate of this decrement is higher at low pressure.

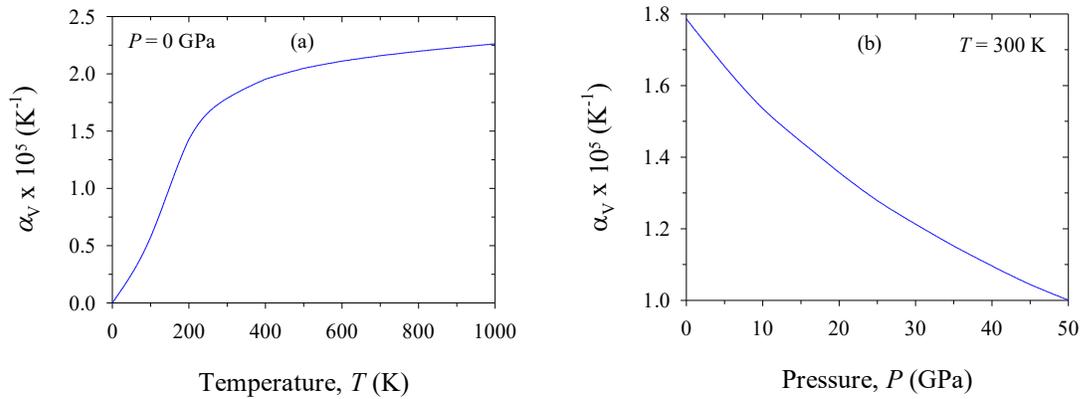

Figure 3: (a) Temperature and (b) Pressure dependences of volume thermal expansion coefficient of NbRuB.

Debye temperature is a fundamental material parameter. From the magnitude of Debye temperature we can estimate a number of important physical properties of solids namely, melting temperature, phonon specific heat, lattice thermal conductivity etc. It is also related to the bonding strength among the atoms present within the crystal. Debye temperature sets the characteristic energy scale for phonon mediated superconductors. The temperature and pressure dependent Debye temperature of NbRuB are shown in Figs. 4a and 4b, respectively.

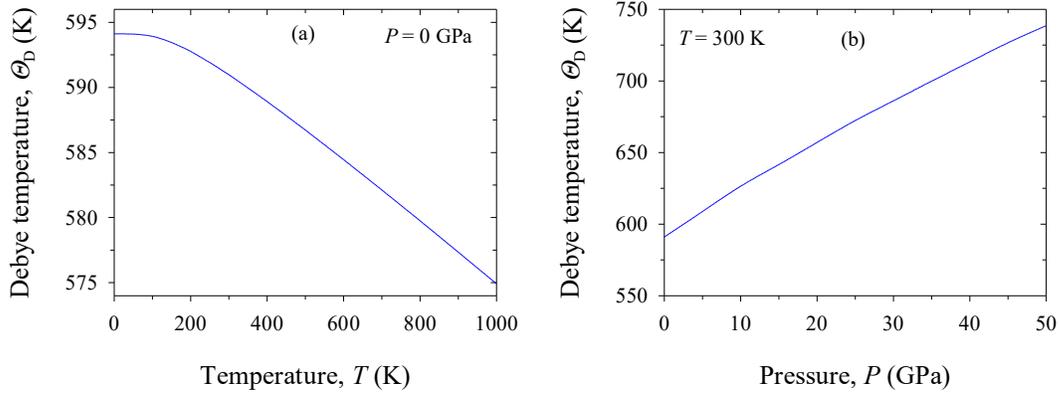

Figure 4: (a) Temperature and (b) Pressure dependences of Debye temperature of NbRuB.

The zero temperature and pressure Debye temperature, $\theta_D$, was found to be 594 K. This reasonably high Debye temperature for NbRuB indicates that this compound should have quite high lattice thermal conductivity. Tian and Chen [30] calculated Debye temperature from the elastic constants and found it to be 578.6 K, close to the value obtained here. Xie et al. [27], on the other hand, obtained $\theta_D$ from the analysis of heat capacity and found it to be 468 K, substantially lower than the one found in this study.

The behavior of materials under different thermodynamical constraints is related to the specific heats of solids. It also determines how efficiently a solid stores heat under different conditions of constant pressure or constant volume. The phonon contribution dominates both the specific heats at constant volume ($C_v$) and constant pressure ($C_p$) at high temperatures. Fig. 5 shows the variation of molar heat capacities of NbRuB with temperature. The low-$T$ specific heat shows a $T^3$ variation in accordance with the Debye model, whereas at high-$T$ both $C_v$ and $C_p$ approaches the classical Dulong-Petit value.

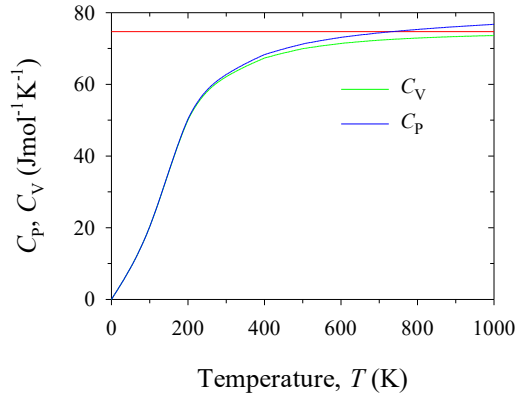

Figure 5: Temperature dependent specific heats of NbRuB. The red line shows the Dulong-Petit limit.

*3.4 Electronic band structure and energy density of states of NbRuB*

The variation of electronic energy state with electron wave vector constitutes the electronic band structure. The study of electronic band structure is essential to comprehend all the electronic transport and optical properties of a material. It also provides us with important information regarding the nature of chemical bonding and structural features. The result of band structure calculations along high-symmetry directions within the *k*-space is shown in Fig. 6. The horizontal dashed line is drawn as the Fermi level. It is seen that several dispersive bands cross the Fermi level, especially along the *Y-S* direction. There is also appreciable band overlap between valence and conduction states. All these imply that NbRuB should exhibit metallic conductivity. The detailed features of band structure and their roles on various electronic phenomena can be explained with the calculated total and partial densities of states. These electronic energy density of states obtained using the GGA functional are shown in Fig. 7.

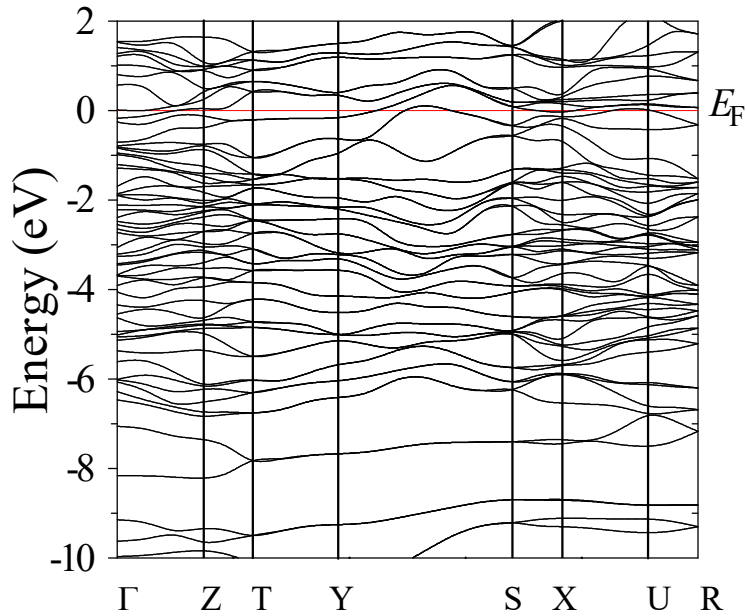

Figure 6: Electronic band structure of NbRuB along the high symmetry directions in the Brillouin zone.

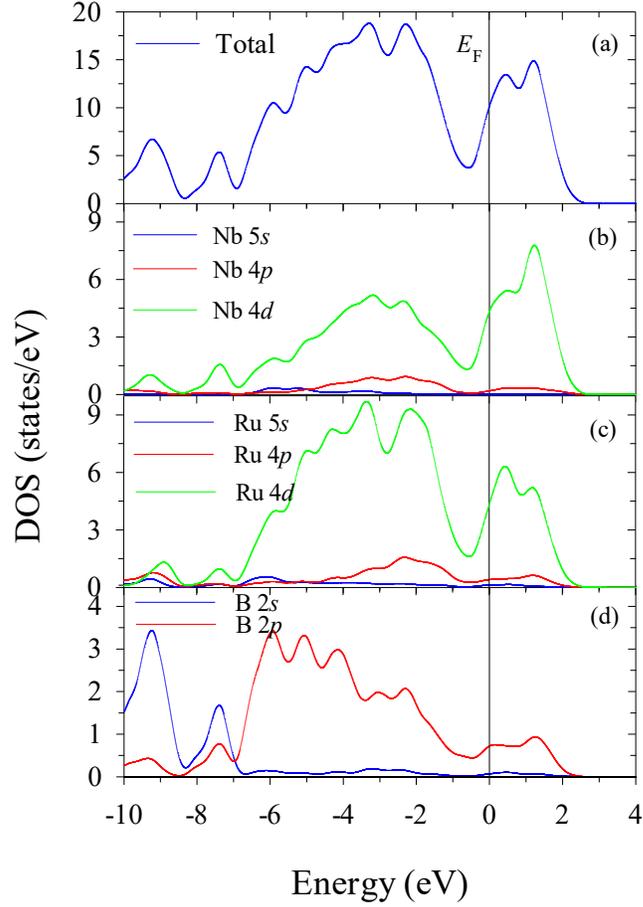

Figure 7: Total and partial electron energy density of states.

The electron energy density of states (DOS) of a system determines the number of quantum states per unit interval of energy at each energy level that are available to be occupied by electrons, normalized to volume or unit cell. The density of states for a given band $n$, $N_n(E)$, is defined in three-dimensions as

$$N_n(E) = \int \frac{d\mathbf{k}}{4\pi^3} \delta(E - E_n(\mathbf{k})) \qquad (2)$$

where $E_n(\mathbf{k})$ describes the dispersion of the given band and the integral is carried over the first Brillouin zone. The total energy density of states, $N(E)$, is obtained by summing over all bands

$$N(E) = \sum N_n(E) = \sum \int \frac{d\mathbf{k}}{4\pi^3} \delta(E - E_n(\mathbf{k})) \qquad (3)$$

Fig. 7a shows a large value of $N(E_F)$, 10.2 states/eV-cell reassuring the metallic character of NbRuB. The atomically resolved partial DOS (PDOS) provides us with more insights regarding the incipient electronic structure. It is seen from Fig. 7 that the low-energy valence band states (below -10 eV to ~ -7 eV) are primarily derived from the B 1$s$ states. The $d$-orbitals of Nb and Ru contribute moderately to low-energy total DOS (TDOS). B 2$p$, Ru 4$d$, and Nb 4$d$ orbitals dominate in the intermediate energy range (from -7 eV to below the Fermi energy). Electronic states located at the levels of and above $E_F$ are delocalized and take part in various transport phenomena. It is seen from Fig. 7 that the TDOS around the Fermi energy arises from almost equal contributions due to 4$d$ electronic orbitals of Nb and Ru atoms. Significant $N(E_F)$ implies that electronic contribution to the heat capacity and thermal conductivity should be appreciable in this transition metal ternary boride. Figs. 7 indicate that there is significant hybridization among the B 2$p$, Nb 4$d$, and Ru 4$d$ orbitals in the vicinity and below the Fermi level. These orbitals participate in the inter-atomic bonding of the crystal. There is a noticeable pseudogap just below the Fermi energy at ~ -0.75 eV. The 2$p$ electronic states of B give rise to strong covalency among B atoms and significantly contribute to the hardness of the material.

## 3.5 Optical properties of NbRuB

Optical properties of a material are closely related to the material's response to incident electromagnetic radiation. The response to visible light is particularly important from the view of optoelectronic applications. The response to the incident radiation is completely determined by the various energy dependent (frequency) optical parameters, namely real and imaginary part of dielectric constants, $\varepsilon_1(\omega)$ and $\varepsilon_2(\omega)$, respectively, real part of refractive index ($n(\omega)$), extinction coefficient ($k(\omega)$), loss function ($L(\omega)$), real and imaginary parts of optical conductivity ($\sigma_1(\omega)$ and $\sigma_2(\omega)$, respectively), reflectivity ($R(\omega)$), and the absorption coefficient ($\alpha(\omega)$). The calculated optical constants of NbRuB for photon energies up to 16 eV with electric field polarization vectors along [100] and [001] directions are shown in Figs. 8. As mentioned earlier, polarization dependent optical parameters yield information regarding optical and electronic anisotropy.

Fig. 8a shows the real and imaginary parts of the dielectric constants. Both $\varepsilon_1(\omega)$ and $\varepsilon_2(\omega)$ show metallic characteristics [44]. For metallic systems, in the low frequency region where $\omega\tau \ll 1$, the imaginary part of the dielectric constant dominates the optical behavior. On the other hand, when $\omega\tau \gg 1$ (high frequency region), the real part approaches unity and the imaginary part becomes very small. Therefore the material becomes transparent to incident electromagnetic radiation. Here, $\tau$ is the electron relaxation time.

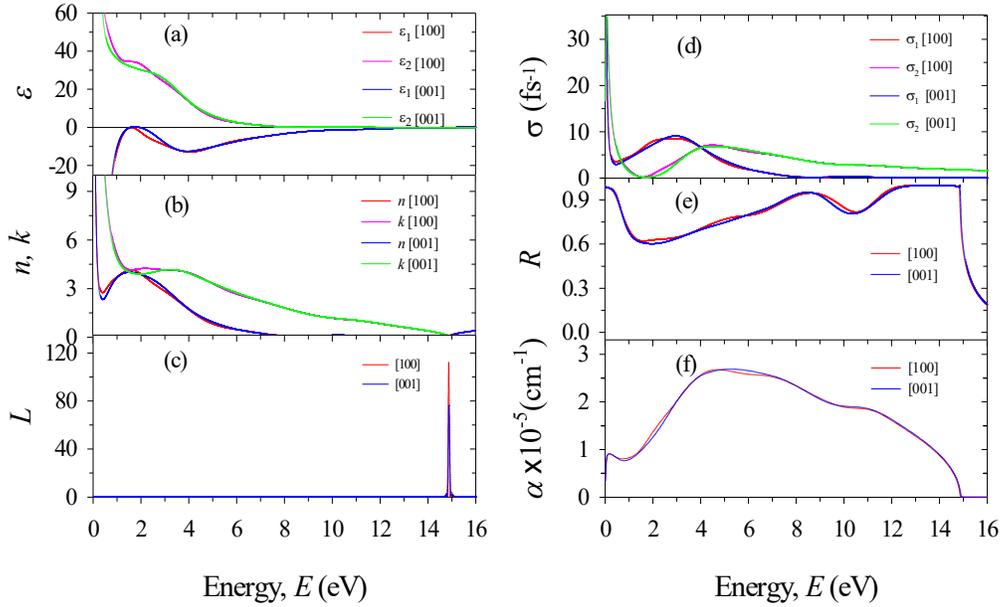

Figure 8: (a) Dielectric constant, (b) Refractive index, (c) Loss function, (d) Optical conductivity, (e) Reflectivity, and (f) Absorption coefficient of NbRuB for different electric polarizations.

The frequency dependence of real and imaginary parts of the refractive indices is given in Fig. 8b. The real part of the refractive index determines the phase velocity of the electromagnetic wave inside the sample, while the imaginary part, often termed as extinction coefficient, gives the measure of the amount attenuation when the electromagnetic wave travels through the material. Both the real and imaginary part of the refractive index are related to the dielectric constant via the relations $\varepsilon_1(\omega) = n^2(\omega) - k^2(\omega)$ and $\varepsilon_2(\omega) = 2n(\omega)k(\omega)$ [45]. Therefore, at high energies both real and imaginary parts of the refractive index becomes small since the imaginary part of the dielectric constant approaches zero, as exhibited in Fig. 8b. Slight polarization dependent anisotropy in the dielectric function is translated into slight anisotropy in the frequency dependent refractive index. This anisotropy, though still low, is relatively pronounced near ~ 6 eV.

The frequency dependent loss function is shown in Fig. 8c. The energy loss function, $L(\omega)$, of a material is an important parameter in the optical property study which is useful for understanding the screened excitation spectra, especially the collective excitations produced by the swift charges traversing a solid. The highest peak of the energy loss spectrum appears at a particular incident light frequency (energy) known as the bulk screened plasma frequency. A sharp electromagnetic loss peak in seen in Fig. 8c. This implies that the underlying excitation spectrum is quite monochromatic. The screened plasma frequency for recently discovered NbRuB is found to be ~ 15 eV. The loss spectrum is quite isotropic with respect to the polarization of the incident radiation.

Optical conductivity spectra are shown in Fig. 8d. The real part of the energy dependent optical conductivity represents the in-phase current which produces the resistive joule heating, while the imaginary part determines the π/2 out-of-phase inductive current. At low frequencies both $\sigma_1(\omega)$ and $\sigma_2(\omega)$ are proportional to the Drude conductivity. It is seen from Fig. 8d that the low-frequency conductivities are high. This shows that NbRuB has good metallic characteristics. The real and imaginary parts of the optical conductivity show broad peaks at ~ 3 eV and ~ 4.5 eV, respectively. Therefore, NbRuB should exhibit significant optical conductivity in the visible range of electromagnetic spectra. In the high-frequency region, above 7 eV, $\sigma_1(\omega) \ll \sigma_2(\omega)$, and the electrons display an essentially inductive character. No energy is absorbed from the electromagnetic field and no joule heat appears. Optical conductivity is found to be fairly isotropic.

The reflectivity is the ratio of the energy of a wave reflected from a surface to the energy of the wave incident on it. The reflectivity spectrum is exhibited in Fig. 8e. $R(\omega)$ rises sharply at around 15 eV, close to the plasma edge. Strong metallic reflection dominates at energies below 15 eV. There are two shallow dips in the reflectivity spectrum centered at ~ 2 eV and ~ 10 eV. The reflectivity spectrum shows almost no polarization dependence. Barring those shallow dips, $R(\omega)$ remains high from infra-red to near ultraviolet energies, without any strong selective characteristic.

Fig. 8(f) illustrates the absorption spectra. The absorption coefficient gives a measure of the distance that a light of a particular wavelength (energy) can penetrate inside a material before it gets absorbed. It also provides us with information about the optimum solar energy conversion efficiency. The absorption spectrum rises sharply above 2 eV, reaches maximum at around 5 eV and decreases to almost zero at and above 15 eV. The broad peak in $\alpha(\omega)$ involves optical transitions of electrons among B 2*p*, Ru 4*d*, and Nb 4*d* electronic orbitals.

## 4. Discussion and conclusions

First-principles DFT based calculations have been performed to investigate the elastic, thermal, electronic, and optical properties of recently discovered transition metal ternary boride NbRuB. The optical properties are investigated for the first time. Some of the thermodynamic properties are also explored theoretically using the quasi-harmonic Debye approximation for the first time. The structural parameters obtained from the optimized geometry within the GGA show excellent agreement with those found by previous experimental and theoretical studies [24, 27, 30]. The estimated elastic constants and moduli show fair correspondence with previous estimates [30]. These elastic constants fulfill Born criteria of mechanical stability. NbRuB shows significant elastic anisotropy. The material is markedly more compressible along the *c*-axis compared to other crystallographic directions. Relatively high values of bulk and shear moduli (298.3 GPa and 155.4 GPa, respectively) of NbRuB shows that it has a potential to be used as a moderately hard material. Pettifor [41] suggested that the Cauchy pressure ($C_{12} - C_{44}$) can be used to describe the character of chemical bonding in solid materials. Cauchy pressure is positive for NbRuB, implying that metallic bonding dominates in this compound. Such materials should exhibit ductile behavior as was found from the calculated values of Poisson's ratio and Pugh's ratio (Table 3). The Poisson's ratio also predicts the character of interatomic forces in solids. A material is said to be a central force solid when Poisson's ratio lies within 0.25 to 0.50, otherwise it is a non-central force solid. Therefore, the structural stability is derived from central forces in NbRuB.

A hard material is able to resist the plastic deformation. It primarily involves preventing the nucleation and movement of dislocations, which lead to irreversible change within the structure. In general, a material with short covalent bonding has a tendency of restricting such motion of dislocations and a material containing more delocalized bonds tolerates them. Accordingly, diamond, the hardest material known to date, have short covalent carbon–carbon bonding that shows high directionality with high strength. On the other hand, metallic compounds consisting of non-directional bonding and are usually soft and ductile due to having characteristic sea of mobile electrons. As an exception to the norm, NbRuB has both ductility and a fair degree of hardness, which can be quite useful for applications.

Pressure and temperature dependent behaviors of the bulk modulus, coefficient volume thermal expansion, and Debye temperature of NbRuB have been studied. A high Debye temperature implies high interatomic force and a significant phonon thermal conductivity. Light mass and strong covalent bonding among B atoms are primarily responsible for this large Debye temperature. NbRuB exhibits conventional BCS superconductivity at 3.1 K. We have calculated the electron-phonon coupling constant, $\lambda_{e\text{-}ph}$, from the $T_c$ equation due to McMillan [46] with estimated Debye temperature of 594 K and an assumed Coulomb pseudopotential, $\mu$, of 0.10. We have found $\lambda_{e\text{-}ph}$ = 0.431, which puts NbRuB in the category of weakly coupled BCS superconductors. It is interesting to note that Os containing ternary borides have stronger chemical bonding [6]. These compounds may have higher Debye temperature and significant electronic density of states at the Fermi level. Such materials may exhibit superconductivity at higher temperature. The specific heat of NbRuB has been calculated as a function of temperature. The low-temperature part follows Debye $T^3$ law, while at high temperatures Dulong-Petit limit is reached. From the calculated value of $N(E_F)$, we have obtained the coefficient of electronic specific heat, $\gamma_e$, given by $\gamma_e$

= $\pi^2 k_B^2 N(E_F)/3$. It shows that the specific heat is primarily dominated by the phonon contribution. At room temperature the electronic contribution is only around 2% of the total heat capacity.

The electronic band structure calculations were performed using both LDA and GGA. No significant difference can be seen. The gross features of calculated electronic band structure agree well with those found in other investigations [27, 30]. The bands crossing the Fermi-level along different symmetry directions within the Brillouin zone are quite dispersive and no significant electronic anisotropy can be detected. The electronic density of states spectrum shows that the Fermi level of NbRuB resides at the rising part of a deep pseudogap (Fig. 7), which indicates that there is a scope to tune $N(E_F)$ to higher or lower values by appropriate chemical substitutions.

The optical constants are studied theoretically for the first time using LDA and GGA functionals. Once again, no significant difference can be detected. All the optical constants were found to be quite isotropic with respect to the polarization of the incident electric field component. The energy dependent optical constants reveal strong metallic characteristics. All these are consistent with the electronic band structure calculations. The reflectivity spectra show some interesting characteristics. Barring two shallow dips, $R(\omega)$ remains high from infra-red to near ultraviolet energies (Fig. 8e). This non-selective nature of $R(\omega)$ can be useful for applications. NbRuB has the potential to be used as an effective reflecting material to reduce solar heating. $R(\omega)$ characteristics of NbRuB are comparable to many MAX compounds in this regard [44, 47-52].

A primary challenge to modern material science is the innovative design and synthesis of new compounds with attractive properties for novel applications. Amongst, search for hard materials is one of the very important subjects. Recently synthesized NbRuB possesses reasonable hardness, ductility, metallic electronic characteristics, non-selective highly reflective optical behavior, and conventional weak-coupling BCS superconductivity. The elastic anisotropy in NbRuB originates from anisotropy in the chemical bonding. On the other hand, electronic and optical characteristics are fairly isotropic. We hope that this theoretical investigation of elastic, thermal, electronic, and optical properties of technologically promising recently discovered NbRuB will stimulate further experimental and theoretical studies in near future.

**Compliance with Ethical Standards**

The authors declare that they have no conflict of interest.